\documentclass[12pt,a4paper]{iopart}

\usepackage{iopams}
\usepackage{graphicx}
\usepackage[breaklinks=true,colorlinks=true,linkcolor=blue,urlcolor=blue,citecolor=blue]{hyperref}
\usepackage{fancyhdr}
\begin{document}

\title{Multi-level cascaded electromagnetically induced transparency in cold atoms using an optical nanofibre interface  }

\author{Ravi Kumar$^{1,2}$, Vandna Gokhroo$^1$ and S\'ile Nic Chormaic$^1$}
\address{$^1$Light-Matter Interactions Unit, Okinawa Institute of Science and Technology Graduate University, Onna, Okinawa 904-0495, Japan \\
$^2$Physics Department, University College Cork, Cork, Ireland}

\ead{sile.nicchormaic@oist.jp}

\begin{abstract}
Ultrathin optical fibres integrated into cold atom setups are proving to be ideal building blocks for atom-photon hybrid quantum networks.  Such optical nanofibres (ONFs) can be used for the demonstration of nonlinear optics and quantum interference phenomena in atomic media.   Here, we report on the observation of multilevel cascaded electromagnetically induced transparency (EIT) using an optical nanofibre to interface cold $^{87}$Rb atoms. Intense evanescent fields can be achieved at ultralow probe (780 nm) and coupling (776 nm) powers when the beams propagate through the nanofibre. The observed multipeak transparency spectra of the probe beam could offer a method for simultaneously slowing down multiple wavelengths in an optical nanofibre or for generating ONF-guided entangled beams, showing the potential of such an atom-nanofibre system for quantum information. We also demonstrate all-optical-switching in the all-fibred system using the obtained EIT effect.

\end{abstract}

\vspace{1pc}

\section{Introduction}
A strong interaction between light and matter is one of the main requirements to successfully realise  neutral atom-based  quantum networks \cite{Cirac1997, Boozer2007, Kimble2008}. Atoms confined in a cavity provide one method of achieving  strong coupling due to the small mode volume of the light field \cite{Miller2005}. However, in order to ensure that photons can travel long distances - as would be desirable in quantum networks - such cavities need to be coupled to optical fibres \cite{Enk1998, Wilk2007, Volz2011, Ritter2012}. Therefore, if strong atom-photon coupling could be achieved in a system that is by itself inherently fibred, this would be a distinct advantage.  Optical nanofibres (ONFs) offer a small optical mode volume in the evanescent field region and are, by default, fibre coupled, thereby providing an alternative method for exploring the strong coupling regime \cite{Kato2015}. Light can be coupled to the pigtails and propagate through the nanofibre waist with extremely high transmission. A large fraction of the `guided' light is contained within the evanescent field, which extends beyond the physical boundary of the fibre \cite{Kien2004}. The evanescent field decays exponentially perpendicular to the light propagation direction, leading to a strong confinement of  light in the transverse direction. Atoms surrounding the waist region couple to the evanescent field \cite{Kien2006, Zoubi2010, Kumar2015b}.  Alternatively, spontaneous emission from the atoms may also couple to the guided modes of the ONF \cite{Klimov2004, Kien2005b, Yalla2012, Liebermeister2014}. This `atom-nanofibre' system could serve as a node in a quantum network, with information being stored in the quantum states of the atoms and information being transferred via the guided light, which acts as the quantum bus. In 2009, Le Kien and Hakuta \cite{Kien2009} proposed that light guided along a nanofibre, passing through a cold atom cloud, could be slowed down.   More recently, such effects have been demonstrated experimentally \cite{Sayrin2015, Gouraud2015}. These demonstrations are based on electromagnetically induced transparency (EIT) in a $\Lambda$-type, three-level atomic system, generating a transparency window in the absorption profile of the probe beam leading to a reduction of the group velocity of the pulse guided by the nanofibre. Here, we use a ladder-type scheme in order to obtain multiple EIT windows, with the potential to support slow group velocities for multiple probe pulses at different frequencies simultaneously. Two light fields propagating with slow group velocities could be used to produce quantum entanglement \cite{Lukin2000}. Furthermore, a ladder type system is generally utilised to study Rydberg atoms \cite{Chang2007}, and multi-level EIT schemes may be useful for nonlinear light generation processes \cite{Wang2010}.

\section{Experimental methods}
\subsection{ONF Fabrication}
A commercial optical fibre is used to fabricate a 350 nm diameter ONF using a heat-and-pull technique. In this method, an unjacketed part of an optical fibre is heated and stretched. There are various ways of heating the fibre including using an oxy-butane or oxy-hydrogen flame, a focussed CO$_2$ laser, microfurnaces, electric strip heaters or sapphire tubes. We use an oxy-hydrogen torch to heat the fibre using a pulling rig which is described elsewhere \cite{Ward2014}. In order to maintain a high optical transmission, the pulling process must be adiabatic. We start with a 250 $\mu$m (cladding) diameter fibre and pull until a waist diameter of $\sim$350 nm is achieved.  The prepared ONF has a transmission of $\sim$84\% for 780 nm light and it remains the same during the experiments. The diameter of the nanofibre ensures that only the fundamental fibre-guided mode, LP$_{01}$, propagates for the wavelengths of light used in the experiments, i.e. 776 and 780 nm. The ONF is installed in a vacuum chamber - used for the magneto-optical trap - in such a way that its waist is aligned with the cloud of cold atoms \cite{Morrissey2009}.

\subsection{Cold atoms}
Atoms are cooled in a standard magneto-optical trap (MOT) to a temperature of $\sim$200 $\mu$K with three cooling beams in the retro-reflected configuration. The cooling beams are kept 14 MHz red-detuned from  the 5 S$_{1/2}$ F=2 $\rightarrow$ 5 P$_{3/2}$ F$^\prime$=3 transition, and the repump beam is kept at the 5 S$_{1/2}$ F=1$\rightarrow$ 5 P$_{3/2}$ F$^\prime$=2 transition. The temperature of the cold atom cloud is measured using a time-of-flight technique by taking the fluorescence image of the cloud after different free expansion times. The cloud is overlapped with the ONF  using two magnetic compensation coils. A small current in the compensation coils is adjusted while monitoring the fluorescence coupling to the ONF via a photon counter connected to one pigtail.  A higher coupling rate indicates better overlap of the densest part of the atomic cloud with the ONF. An optical depth of 0.17 is achieved when the probe beam passes through the ONF during continuous operation of the MOT. Note that there are approximately two atoms in the evanescent field region of the ONF contributing to the observed signal.

\begin{figure}[ht]
\centering
\includegraphics[width=7.5 cm]{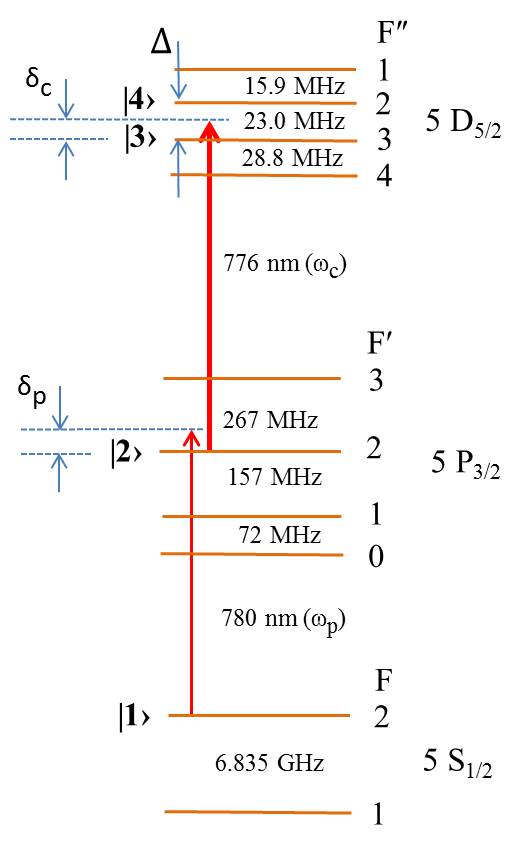}
\caption{ Energy level diagram for $^{87}$Rb. The relevant levels for the EIT experiment are marked as $\vert 1 \rangle $, $\vert 2 \rangle $, $\vert 3 \rangle $ and $\vert 4 \rangle $. }
\label{fig:energylevel}
\end{figure}
\begin{figure}[ht]
\centering
\includegraphics[width=10 cm]{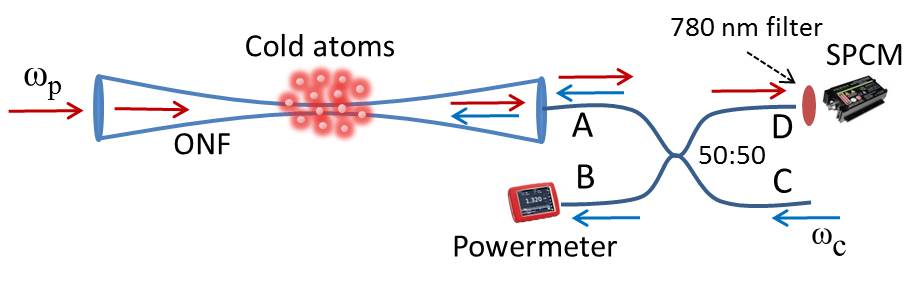}
\caption{Schematic of the experimental setup. ONF: Optical nanofibre, SPCM: Single photon counting module, 50:50:fibre beam splitter with 50:50 split ratio. One port (port A) of the fibre splitter is spliced to one pigtail of the tapered fibre. }
\label{fig:Schematic}
\end{figure}

\subsection{Ladder type EIT system}
We consider a multilevel, cascaded EIT system as shown in Fig.\ref{fig:energylevel}. An intense coupling beam at 776 nm ($\omega_c$) drives the upper transition, 5 P$_{3/2} \rightarrow$ 5 D$_{5/2}$,  and a weak probe beam ($\omega_p$) at 780 nm drives the lower transition, 5 S$_{1/2} \rightarrow$ 5 P$_{3/2}$. Both the probe and coupling beams are sent through the optical nanofibre in a counter-propagating configuration (Fig. \ref{fig:Schematic}). A power meter connected to port B of a 50:50 fibre beam splitter measures the power level of the coupling beam  sent into the tapered fibre pigtail connected to port A. The probe beam propagates in the opposite direction to the coupling beam and half of the probe power is obtained at port D, where it is out-coupled and passed through a 780 nm filter (FWHM: 3 nm) in free space and then directed to a single photon counter (SPCM). The filter prevents the 776 nm back-reflected light from reaching the detector.

\section{Results and discussions}
EIT reduces the absorption of a weak probe beam in resonance with a dipole-allowed atomic transition due to the presence of a strong coupling beam on a linked transition. Optical properties of atomic samples are mainly associated with the energy levels and the imaginary part of the susceptibility,  Im[$\chi^{(1)}$], determines the absorption of a probe beam passing through the atomic gas.  The susceptibility is modified by the coupling beam which can be analysed by the the time evolution of the density matrix of the system \cite{Gea-Banacloche1995, Anil2009, Doai2014}. Considering the multi-level system as shown in Fig. \ref{fig:energylevel}, the absorption coefficient for the probe beam, using the dipole- and rotating wave-approximations, can be given as \cite{Doai2014},
\begin{equation}
\alpha \propto \omega_p \frac{B}{A^2 + B^2}
\end{equation}
with \begin{equation*}
A = - \delta_p + \frac{A_{32}}{\gamma_{32}} +  \frac{A_{42}}{\gamma_{42}} , B = \gamma_{21} + \frac{A_{32}}{\delta_p + \delta_c} + \frac{A_{42}}{\delta_p + \delta_c -\Delta} ,
\end{equation*}
\begin{equation*}
A_{32}= \frac{\gamma_{31}(\delta_p + \delta_c)}{\gamma_{31}^2 + (\delta_p + \delta_c)^2} a_{32}^2 (\frac{\Omega_c}{2})^2 , A_{42}= \frac{\gamma_{41}(\delta_p + \delta_c -\Delta)}{\gamma_{41}^2 + (\delta_p + \delta_c -\Delta)^2} a_{42}^2 (\frac{\Omega_c}{2})^2
\end{equation*}
 where $\omega_p$ is the frequency of the probe beam, $\Omega_c$ is the Rabi frequency of the coupling beam, $\delta_p$ ($\delta_c$) is the detuning of the probe (coupling) beam, $\Delta$ is the frequency difference between $\vert 3\rangle$ and $\vert 4\rangle$, $\gamma_{ij}$ are the decay rates from $\vert i\rangle$ to $\vert j\rangle$ and $a_{32}$ ($a_{42}$) is the relative transition strength between levels $\vert 2\rangle$ to $\vert 3\rangle$ ($\vert 4\rangle$). Here, we have neglected the transition 5 P$_{3/2}$ F$^\prime$=2 $\rightarrow$5 D$_{5/2}$ F$^{\prime \prime}$=1 as this is weak and not observed in our experiments.

\subsection{Multilevel EIT}

We study multilevel EIT using the hyperfine levels of three fine structure levels, namely 5 S$_{1/2}$, 5 P$_{3/2}$, and 5 D$_{5/2}$ (Fig. \ref{fig:energylevel}). $\omega_c$ is kept at a fixed blue-detuning ( $\delta _c$ = 7 MHz) from 5 P$_{3/2}$ F$^{\prime}$=2 $\rightarrow$ 5 D$_{5/2}$ F$^{\prime \prime}$=3, whereas $\omega _{p}$ is scanned across 5 S$_{1/2}$ F=2 $\rightarrow$ 5 P$_{3/2}$ F$^{\prime}$=2. In fact, in this situation the coupling beam is coupled to all hyperfine levels of 5 D$_{5/2}$ simultaneously with different detunings. While scanning the probe, EIT dips appear in the probe transmission when a resonant condition is met with any of the hyperfine transitions of the 5 D$_{5/2}$ level. The hyperfine levels in 5 D$_{5/2}$ are closer spaced than the hyperfine levels of 5 P$_{3/2}$. The probe scan does not cover any hyperfine levels of 5 P$_{3/2}$ other than $F^{\prime}=2$; however, it covers all the hyperfine levels of 5 D$_{5/2}$. The frequency reference is obtained using a vapour cell with a portion of the probe and  coupling beams sent through it in a counter-propagating configuration while simultaneously monitoring the probe transmission in the vapour cell and through the nanofibre in the cold atoms experiment.

\begin{figure}[ht]
\centering
\includegraphics[width=10cm]{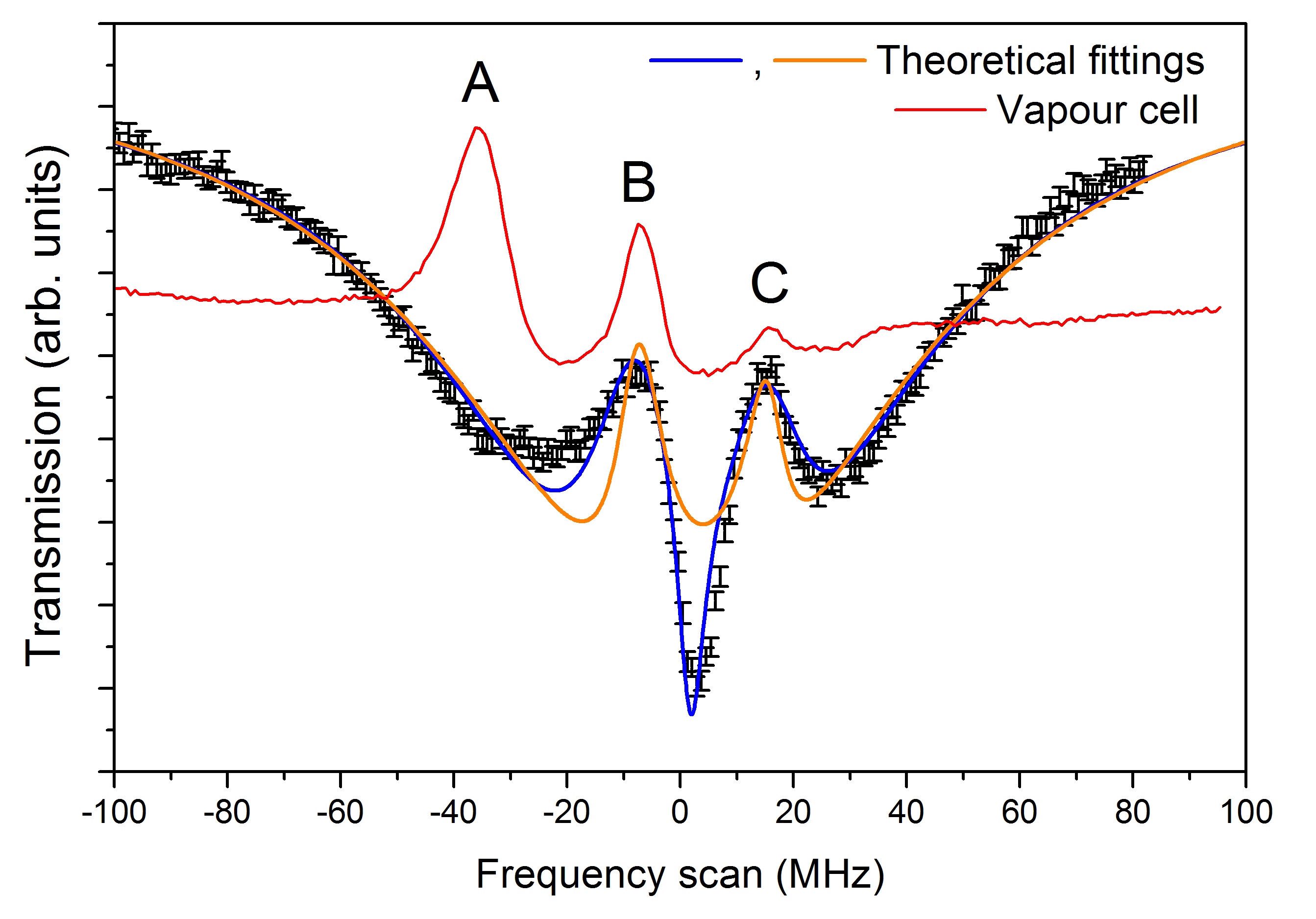}
\caption{Transmission through the fibre as the frequency  of the probe beam is scanned. The coupling beam is 7 MHz blue-detuned from the transition 5 P$_{3/2}$ F$^\prime$=2 $\rightarrow$ 5 D$_{5/2}$ F$^{\prime \prime}$=3 and has a power of 200 nW. Black points: experimental data averaged for 200 runs. Orange curve: fitting using equation 1. Blue curve: fitting with a Lorentzian curve added to equation (1). Red curve: frequency reference obtained by a vapour cell. In the vapour cell, some atoms can make the transitions involving 5 P$_{3/2}$ F$^\prime$=3 due to Doppler shift. Peak A corresponds to the transition 5 P$_{3/2}$ F$^\prime$=3 $\rightarrow$ 5 D$_{5/2}$ F$^{\prime \prime}$=4, peak B due to 5 P$_{3/2}$ F$^\prime$=3 $\rightarrow$ 5 D$_{5/2}$ F$^{\prime \prime}$=3 and 5 P$_{3/2}$ F$^\prime$=2 $\rightarrow$ 5 D$_{5/2}$ F$^{\prime \prime}$=3, whereas, peak C is due to 5 P$_{3/2}$ F$^\prime$=3 $\rightarrow$ 5 D$_{5/2}$ F$^{\prime \prime}$=2, 5 P$_{3/2}$ F$^\prime$=2 $\rightarrow$ 5 D$_{5/2}$ F$^{\prime \prime}$=2 and 5 P$_{3/2}$ F$^\prime$=1 $\rightarrow$ 5 D$_{5/2}$ F$^{\prime \prime}$=2. Peaks B and C have an equivalence in cold atoms as 5 P$_{3/2}$ F$^\prime$=2 $\rightarrow$ 5 D$_{5/2}$ F$^{\prime \prime}$=3 and 5 P$_{3/2}$ F$^\prime$=2 $\rightarrow$ 5 D$_{5/2}$ F$^{\prime \prime}$=2, respectively, but there is no peak corresponding to peak A since the atomic velocities are not sufficient to allow the 5 P$_{3/2}$ F$^\prime$=3 transition to be within the range of the probe scan. The 5 P$_{3/2}$ F$^\prime$=2 $\rightarrow$ 5 D$_{5/2}$ F$^{\prime \prime}$=1 transition is weak and not observable in either case.  }

\label{fig:comparision}
\end{figure}

A typical EIT profile obtained in the cold atoms with 5 pW of power ($P_p$) in the probe beam  and 200 nW of power ($P_c$) in the coupling beam through the ONF is shown in Fig. \ref{fig:comparision}. Fitting cold atom data with equation 1 (orange curve in Fig. \ref{fig:comparision}) we find that the curve does not fit the central part completely, but provides a very good fit for detunings higher than 25 MHz or lower than $-$25 MHz. Equation 1 is for free beam experiments where it is assumed that all the atoms contributing to the signal experience the same probe and coupling powers which is not true for our experiment. Symmetry of the evanescent field around the tapered region depends on the polarization of light \cite{Kien2004}. In principle, linear polarization does not give a symmetrical field intensity around a nanofibre. In our experiment, the evanescent field  of the coupling beam may not be overlapping with the  probe due to their different polarization orientations at the nanofibre waist. Hence, it is possible that some atoms may observe more 780 nm light, leading to single photon absorption of a portion of the 780 nm probe beam where the 776 nm presence is minimal due to azimuthal field dependence. This may introduce an additional absorption at the 5 S$_{1/2}$ F=2 $\rightarrow$ 5 P$_{3/2}$ F$^{\prime}$=2 resonance with a Lorentzian profile. Hence, we add a Lorentz curve with a natural linewidth to equation 1 and get a reasonably good fit to the experimental data (blue curve in Fig. \ref{fig:comparision}).

\begin{figure}[ht]
\centering
\includegraphics[width=\linewidth]{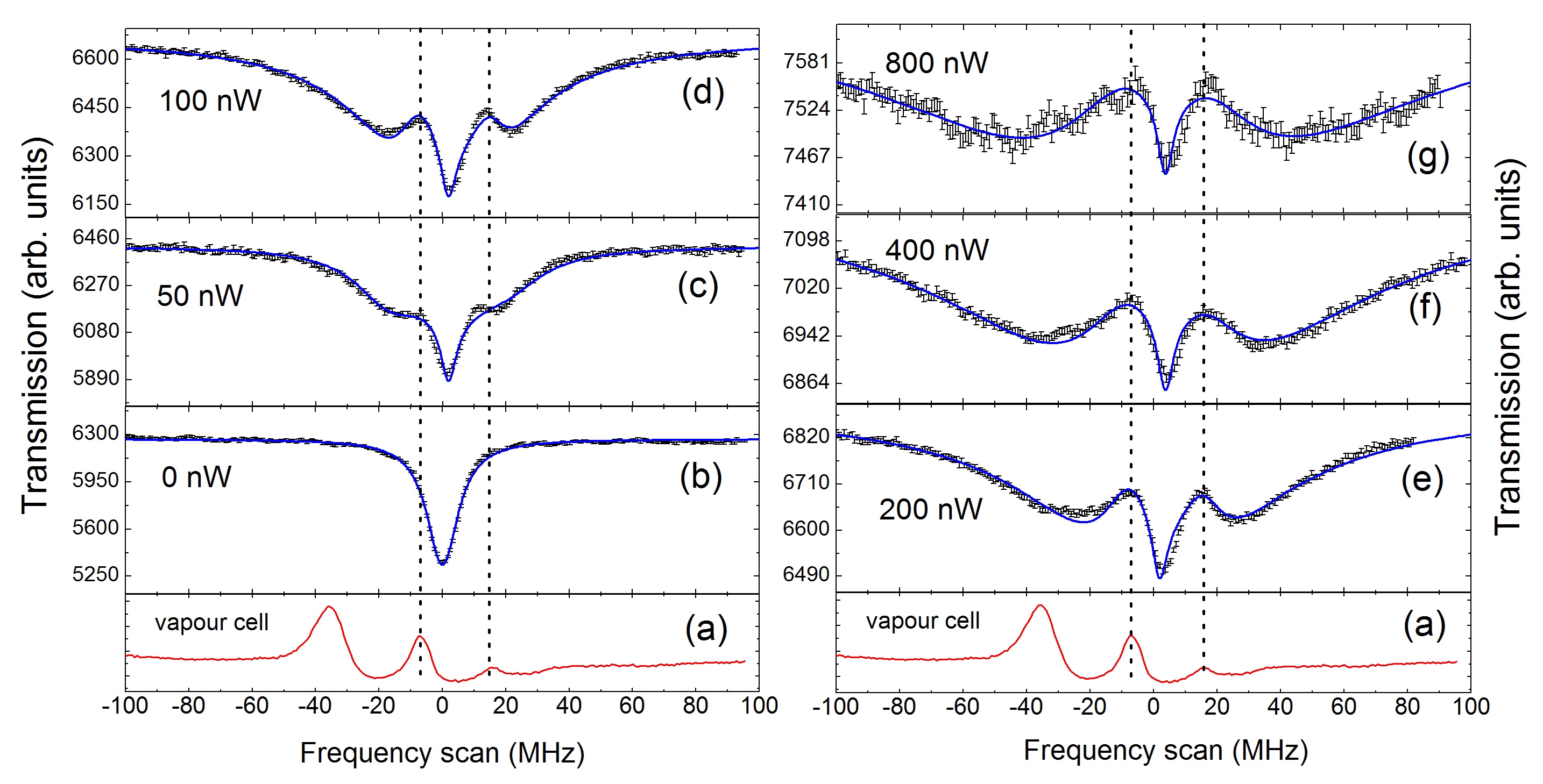}
\caption{Multiple EIT peaks for coupling beam powers from 0 $-$ 800 nW as indicated on the graphs. (a) The frequency reference signal obtained in a vapour cell (red curves) is used for  frequency calibration. (b) At $P_c=0$ nW an absorption signal is observed corresponding to the probe power of 5 pW. (c)-(g) EIT signals for $P_c=50-800$ nW.  The vertical dotted lines show the positions corresponding to the 5 P$_{3/2}$ F$^{\prime}$=2 $\rightarrow$ 5D$_{5/2}$ F$^{\prime \prime}$=3 and F$^{\prime \prime}$=2 transitions. Black dots are experimental data and the blue curves are theoretical fittings. }
\label{fig:EIT}
\end{figure}
\begin{figure}[ht]
\centering
\includegraphics[width=\linewidth]{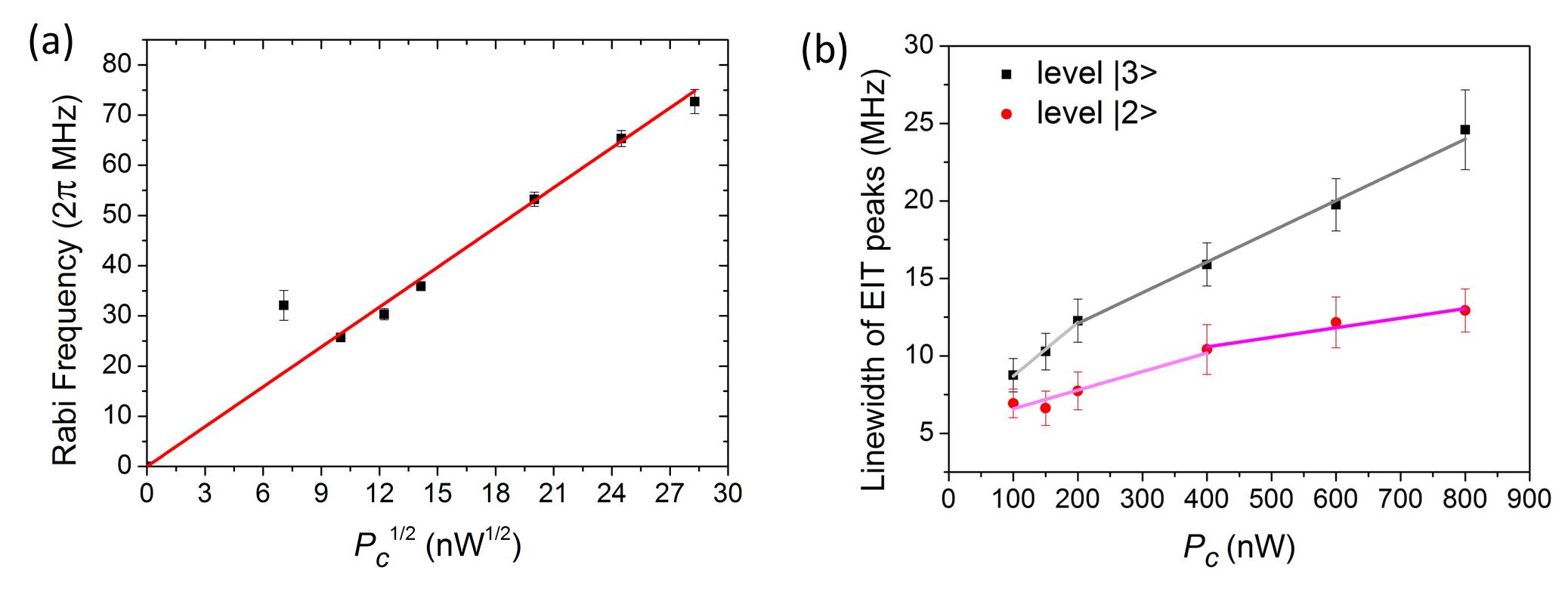}
\caption{(a) Rabi frequency as a function of the square root of power in the coupling beam, determined from Fig. \ref{fig:EIT} for the strong coupling beam condition. The red line is a linear fit to the experimental data. (b) Linewidth of the two transparency peaks as a function of coupling power. The solid straight lines represent linear fits to the experimental data and two clear slopes are evident for each peak. }
\label{fig:RabiFreqNLinewidth}
\end{figure}

 Parameters  $a_{32}$ and $a_{42}$ in equation 1 are the relative transition strengths, which depend on the pump and probe polarizations and the distribution of atoms in magnetic sublevels. In reality, the polarization pattern at the waist is complicated and cannot be taken simply as linearly polarization even for linearly polarized light inputs, as in our case. In fact, the light does not even remain  transverse, but acquires a longitudinal component at the waist region \cite{Kien2004}. Also, aside from the birefringence of the ONF, the pigtails are not polarization maintaining. Hence, these values must be determined from the experimental data. We fit the data for various coupling powers using $a_{32}$ and $a_{42}$ as free parameters and find that these values are almost constant with $a_{32} =0.747$ and $a_{42}=0.624$ with standard deviations of 0.036 and 0.124, respectively. Subsequently, we fit all the data while keeping these values fixed. The obtained fittings are shown in Fig. \ref{fig:EIT} for various coupling powers. The Rabi frequency is kept as a free parameter and we find that it varies linearly with the square root of the coupling power as expected (Fig. \ref{fig:RabiFreqNLinewidth}a).

  The measured linewidths of the obtained peaks with respect to the coupling power are shown in Fig. \ref{fig:RabiFreqNLinewidth}b. There are two noticeable features: (i) broad linewidths, particularly at high $P_c$, compared to the natural linewidth of 6 MHz, and (ii) a linear increase in the width `in sections' as a function of $P_c$. The broadness of the peaks may be due to several reasons. It is well known that the ratio of $P_p/P_c$ and their polarizations affect the EIT profile significantly \cite{Moon2005}. In any studies with an ONF, the atoms are distributed in the exponentially decaying probe and coupling field intensities. Therefore, any particular probe and coupling power does not mean that all the participating atoms observe the same probe and coupling intensities. In other words, an effective $P_p/P_c$, as viewed by any particular atom, has a wide range of values. The polarization of the two beams as observed by the atoms is also complex giving rise to the broadening of the peaks.
Decoherence due to atoms hitting the ONF may also play a contributing role. The width of the EIT peaks increases linearly with $P_c$ to a certain level and then a change in slope is observed for both the peaks, but at different power levels. The origin of this change in slope is yet to be understood via a full theoretical treatment of the problem, but is considered beyond the scope of this current work.  

\subsection{All-fibred-all-optical-switching}

\begin{figure}[ht]
\centering
\includegraphics[width=10 cm]{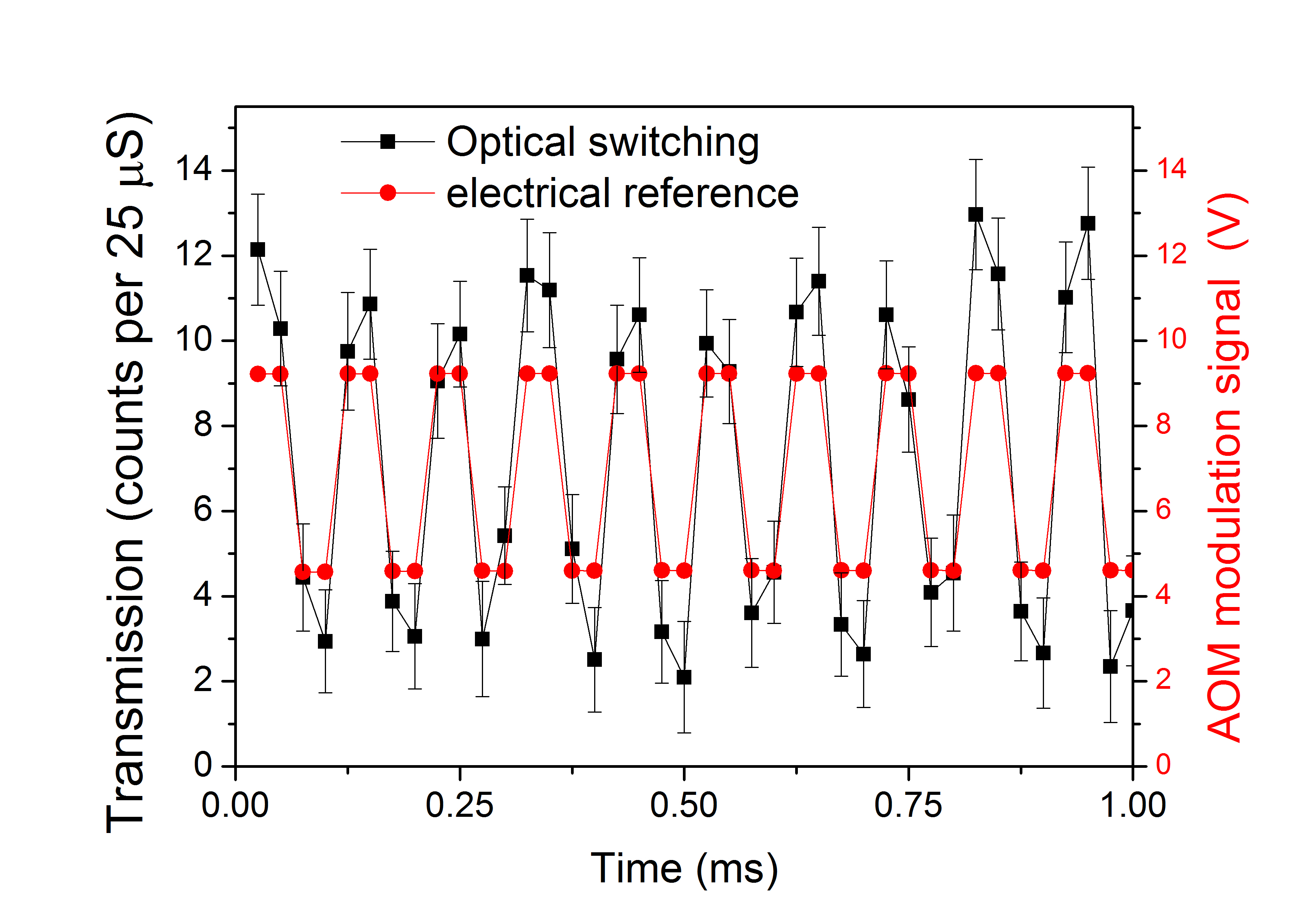}
\caption{Transmission through nanofibre as a function of time to demonstrate all-optical-switching. Black: Photon counts obtained with 10 kHz on-off modulation of the coupling beam in the presence of cold atoms with background counts subtracted. Red: electrical reference signal used for switching  on and off the coupling beam. The data is averaged for 150 runs.}
\label{fig:OpticalSwitch}
\end{figure}

The observed EIT effects in the system are used for making an all-optical-switch. The probe beam is locked to 5 S$_{1/2}$ F=2 $\rightarrow$ 5 P$_{3/2}$ F$^{\prime}$=2 and the coupling beam to  5 P$_{3/2}$ F$^{\prime}$=2 $\rightarrow$ 5D$_{5/2}$ F$^{\prime \prime}$=2. The coupling beam is passed through two AOMs successively, one in ``$+$1'' order and other in ``$-$1'' order with the same frequency, in order to have the facility to switch on and off the beam, $P_c$, without any effective frequency shift. The coupling beam power is set to 80 nW and it is switched off and on at 10 kHz by modulating the AOM. The probe beam is continuously on and the transmission through the nanofibre is monitored. The transmission is higher when the coupling beam is on and lower when off. There is some leakage of the coupling beam due to back reflections on the detector even with the 780 nm filter in place. This back refection is determined by performing the same on and off switching sequence of the coupling beam in the absence of the cold atom cloud and is subtracted from the signal obtained when the cloud is present to get the change in transmission purely due to atoms as a switching medium. Fig. \ref{fig:OpticalSwitch} shows the background-subtracted data and the electrical signal used to control the AOM modulation to switch the coupling beam on and off. One data point is collected every 25 $\mu$s. The switching speed could go higher with a denser cloud, making the data collection gate time even shorter.

\section{Conclusion}
We have demonstrated multi-level EIT in a cascaded system using an ONF for sending the probe and the coupling beams to a cold atomic cloud. This method could be used for storing two light pulses of different frequencies at the same time, and could also be used for generating entanglement with two different frequency beams. Using the EIT results, we have also demonstrated an all-fibred-all-optical switch which could be used for optical data processing in quantum systems. For analysing the EIT peaks, we have neglected  optical pumping effects, such as double resonant optical pumping (DROP) \cite{Moon2007}, which is due to changes in the population of the ground level being probed. This can be justified as the probe beam we use is very weak and the MOT beams are continuously on during the experiment. Due to the presence of the cooling and the repump beams coupling to the two ground states (5 S$_{1/2}$ F=2 and 5 S$_{1/2}$ F=1), the population of these two states should be constant. Reduction of the DROP effect due to the presence of the repump beam has been studied for Rb vapour recently \cite{Ali2015}. The width of the EIT signal is dependent on the coherence properties of atoms in the system. In our system, the atoms are moving in the MOT and the observation region is limited to the evanescent field region of the ONF.  This can be transited by the atoms in a few 10s of $\mu$s. Also, the MOT is in continuous operation, hence the magnetic field and the MOT beams are on during the measurement leading to  further boarding of the obtained signals. However, we have obtained well-separated peaks (23 MHz apart) corresponding to the hyperfine transitions of the 5 D$_{5/2}$ level.  If the atoms were in an ONF-based trap \cite{Goban2012}, the linewidths could be significantly lower and this would be a benefit for slowing down the light beams travelling through the ONF. Also, the signal strength could be higher as the number of atoms in the evanescent field region could be orders of magnitude higher than in the MOT case. This would be an advantage for all-optical-switching to obtain a higher contrast between the on and off states and could also lead to higher speed operation of the switch.

During the preparation of this manuscript we became aware of similar work done using a warm atomic vapour, in which controlled polarization rotation has been demonstrated \cite{Jones2015}.

\section*{Acknowledgements}
We thank Jordi Mompart for useful discussions. This work was supported by the Okinawa Institute of Science and Technology Graduate University. S.N.C. is grateful to JSPS for partial support from Grant-in-Aid for Scientific Research (Grant No. 26400422).

\end{document}